\documentclass[a4paper,fleqn,usenatbib]{mnras}

\usepackage[T1]{fontenc}
\usepackage{ae,aecompl}

\usepackage{graphicx}	
\usepackage{amsmath}	
\usepackage{amssymb}	

\title[Rotating twisted magnetosphere of magnetars]{Rotating twisted magnetosphere of magnetars: approximate analytical solutions}

\author[Tong \& Chen]{H. Tong$^{1}$\thanks{E-mail: tonghao@gzhu.edu.cn},
L. Chen$^{2}$
\\
$^{1}$School of Physics and Materials Science, Guangzhou University, Guangzhou 510006, China\\
$^{2}$Shanghai Astronomical Observatory, Chinese Academy of Sciences, 80 Nandan Road, Shanghai 200030, China
}

\date{Accepted XXX. Received YYY; in original form ZZZ}

\pubyear{2015}

\begin{document}
\label{firstpage}
\pagerange{\pageref{firstpage}--\pageref{lastpage}}
\maketitle

\begin{abstract}
An approximate analytical solution for the rotating twisted magnetosphere of magnetars is presented. The poloidal flux is approximated by the self-similar twisted dipole field. The toroidal field is obtained by the minimum torque model. Under this approximation, it is found that: (1) The Y-point radius decreases with the increase of twist of the magnetic field. (2) The polar cap is larger for larger twist. (3) The particle outflow luminosity is larger for larger twist. (4) The maximum acceleration potential, pulse width of magnetar radio emission both increase with the twist. (5) For an untwisting magnetosphere, the physical properties evolve toward that of the normal pulsars. The above findings are consistent with previous analytical and numerical results. The larger polar cap may correspond to the hot spot during magnetar outburst. In general, a rotating twisted magnetosphere has larger open field line regions. The radio emission of magnetars and fast radio bursts may both originate in the larger and evolving open field line regions of magnetars.
\end{abstract}




\section{Introduction}

The magnetosphere is the basis for pulsar multiwave emissions and spin-down torque. The change and evolution of the magnetosphere will also affect both the timing and pulse profile of pulsars (Kramer et al. 2006; Lyne et al. 2010; Desvignes et al. 2019). Traditionally, a dipole field is often assumed in the study of pulsar electrodynamics and emission theories (Goldreich \& Julian 1969; Ruderman \& Sutherland 1975; Arons \& Scharlemann 1979; Cheng et al. 1986; Qiao et al 2004). Up to now, a dipole field is assumed in some studies of pulsar pulse profile and timing (Gangadhara 2004; Zhang et al. 2007; Cao et al. 2024; Roy et al. 2025).

A self-consistent study of pulsar magnetosphere structure may be based on the ``pulsar equation" (Mestel 1973; Michel 1973; Scharlemann \& Wagoner 1973). Numerical solutions are obtained at the end of the 20th century (Contopoulos et al. 1999). There are many later developments (Gruzinov 2005; Timokhin 2006; Spitkovsky 2006; Philippov et al. 2015; Stefanou et al. 2023; see Contopoulos 2024a,b and references therein for an overview of previous works). The gamma-ray and radio emissions of  pulsars may be model based on this physical magnetosphere (Kalapotharakos et al. 2014; Philippov et al. 2020; see Philippov \& Kramer 2022 for review).

Compared with normal pulsars, magnetars may be neutron stars powered by their strong magnetic field (Duncan \& Thompson 1992). The magnetic field in the case of magnetars carries free energy because the magnetic field may be twisted or have more complex geometry (Thompson et al. 2002; Pavan et al. 2009; Beloborodov 2009; Glampedakis et al. 2014; Akgun et al. 2016; Kojima et al. 2017; Tong 2019). The evolving magnetosphere of magnetars and/or crustal field is responsible for their bursts and outbursts (Coti Zelati et al. 2018; Tong \& Huang 2020). Since magnetars typically have a rotation period about 10 seconds, some modelings of magnetar magnetosphere assume that all the field lines are closed field lines and there is no open field line in the case of magnetars. This will make the magnetosphere of magnetars very different from that of pulsars. However, observationally some high magnetic field pulsars also show magnetar activities (Gavriil et al. 2008; Gogus et al. 2016; Archibald et al. 2016). The radio emission  properties of magnetars are different from that of normal pulsars. However, at the end of their radio activities, they evolve from typical magnetar case to that of the normal pulsar case (Camilo et al. 2016; Huang et al. 2023). And it is a general belief that pulsars and magnetars belong to a unified population of neutron stars (Rea et al. 2010; Vigano et al. 2013).

The twisted magnetic field of magnetars may be described in a self-similar way: $B \propto r^{-(2+n)}$ (Thompson et al. 2002; Pavan et al. 2009; Tong 2019). Due to the twist of magnetic field, the field lines will inflate in the radial direction (Wolfson 1995; Thompson et al. 2002). Noting this point, by inducing the light cylinder radius, it is found that magnetars may have large polar caps (Tong 2019). Of course, the introduction of light cylinder radius is not a self-consistent treatment of the pulsar equation. Recently, numerical simulations of twisted magnetar magnetosphere are reported (Ntotikas et al. 2024, 2025). The authors found that for a twisted magnetosphere, the Y-point radius can be much smaller than the light cylinder radius. Consequently, the polar cap radius and spindown luminosity are much larger than the dipole case. Inspired by the numerical progress, we try to provide an approximate analytical solution of the rotating twisted magnetosphere of magnetars.

The existence of large polar cap and open field line regions (Tong 2019; Ntotikas et al. 2024, 2025) are not only related to the magnetosphere physics of magnetars and the unification of pulsars and magnetars. It  may also be related to the physics of fast radio bursts. The observations of magnetar SGR 1935+2154 tell us magnetars may host fast radio bursts (CHIME/FRB Collaboration 2020; Bochenek et al. 2020). If there are large open field line regions in the case of magnetars, then the fast radio bursts can be generated in the open field line regions. Then it is natural that they share similar physics to that of pulsars and magnetars, e.g., rotating vector model (Mckinven et al. 2025; Liu et al. 2025), polarization (Luo et al. 2020), radius to frequency mapping (Wang et al. 2019; Lyutikov 2020; Tong 2022). Radio wave propagating in the open field line regions may also be easier to break out the magnetosphere (Beloborodov 2021; Qu et al. 2022).

The structure of this paper is: In Section 2, the approximate analytical solution is provided. Its comparison with the numerical simulations is presented in Section 3. Its application to magnetar outburst, especially magnetar radio emissions, is discussed in Section 4. The conclusion is presented in Section 5.

\section{Approximate analytical solutions}

For an axisymmetric magnetosphere, the magnetic field may be describe by the poloidal flux function $\Psi$ (which is related to the poloidal magnetic field) and poloidal current $\Phi$ (which is related to the toroidal magnetic field), following the symbol system in Chen \& Zhang (2021), in spherical coordinate $(r, \theta, \phi)$, the magnetic field is:
\begin{equation}\label{eqn_B}
  {\bf B} = \frac{1}{r \sin\theta} \left[ \frac{1}{r} \frac{\partial \Psi}{\partial \theta} \hat{r} - \frac{\partial \Psi}{\partial r} \hat{\theta} + \Phi \hat{\phi} \right].
\end{equation}
The force-free condition result in a partial differential equation for the flux and current, the so-called ``pulsar equation" (Scharlemann \& Wagoner 1973):
\begin{equation}
  \frac{\partial^2 \Psi}{\partial r^2} + \frac{1}{r^2} \frac{\partial^2 \Psi}{\partial \theta^2} - \frac{\cot\theta}{r^2} \frac{\partial \Psi}{\partial \theta}
\end{equation}
\begin{equation}
   + \Phi^\prime \Phi -\left[ \frac{\partial^2 \Psi}{\partial r^2} + \frac{2}{r} \frac{\partial \Psi}{\partial r} + \frac{1}{r^2} \frac{\partial^2 \Psi}{\partial \theta^2} + \frac{\cot\theta}{r^2} \frac{\partial \Psi}{\partial \theta}\right](\Omega r \sin\theta)^2 =0, \nonumber
\end{equation}
where $\Omega$ is the angular velocity of the compact star (assumed to be a constant), $\Phi^\prime \equiv d\Phi/d\Psi$. The first line in the above equation is named as the ``non-rotational term", while the second line is named as the ``rotational term"\footnote{In the closed field line regions, the toroidal field is due to twist of the magnetic field. While, in the open field line regions, the current and toroidal field are due to the rotation of the neutron star, see eq.(\ref{eqn_Phi_main}). In the following, we are mainly interested in the open field line regions or the last open field line. Therefore, the $\Phi^\prime \Phi$ term is grouped into the ``rotational term".} (Chen \& Zhang 2021). Actually this pulsar equation finds its application in both pulsars and black holes\footnote{In the black hole case, the angular velocity may not be a constant, e.g., for field lines anchered on the accretion disk. The pulsar equation will also contain additional terms which depends on $\Omega^\prime$.} (See Narayan et al. 2007 and references therein.). For a nonrotating magnetar $\Omega=0$, as have been considered by some previous works, the pulsar equation reduces to the case of Grad-Shafranov equation (Wolfson 1995; Thompson et al. 2002; Pavan 2009; Tong 2019):
\begin{equation}
  \frac{\partial^2 \Psi}{\partial r^2} + \frac{1}{r^2} \frac{\partial^2 \Psi}{\partial \theta^2} - \frac{\cot\theta}{r^2} \frac{\partial \Psi}{\partial \theta}  + \Phi^\prime \Phi=0.
\end{equation}
The pulsar equation and the Grad-Shafranov equation can be solved numerically. At the same time, there also exist some approximate analytical solutions. These analytical solutions demonstrate the physics involved more easily.

When studying the analytical solution of pulsar equation for black hole jets and winds, it is found that (Narayan et al. 2007; Tchekhovskoy et al. 2008; Chen \& Zhang 2021), (1) both the ``non-rotation term" and the ``rotation term" can be can be viewed as the approximate solution of the full pulsar equation. (2) The poloidal current is related to the flux function by the minimum torque model (Michel 1969; Narayan et al. 2007; Tchekhovskoy et al. 2008; Chen \& Zhang 2021):
\begin{equation}\label{eqn_Phi_main}
  \Phi = -2 \Omega \Psi = -2 \frac{\Psi}{R_{\rm lc}}.
\end{equation}
Please note that in the above equation (eq.(A44) in Tchekhovskoy et al. 2008; eq.(35) in Chen \& Zhang 2021), the speed of light is set to be $c=1$. Therefore, the light cylinder radius is $R_{\rm lc} = 1/\Omega$. Motivated by this approximation, for the case of rotating twisted magnetosphere, we already know the self-similar analytical solution (Wolfson 1995; Thompson et al. 2002; Pavan et al. 2009; Tong 2019):
\begin{equation}\label{eqn_Psi_main}
  \Psi = r^{-n} f(\theta),
\end{equation}
where $0\le n \le 1$ is a parameter characterizing the twist of the magnetic field. Substituting eq.(\ref{eqn_Psi_main}) into eq.(\ref{eqn_B}), it can be seen that (Wolfson 1995), the magnetic field depends on the radius as $B\propto r^{-(2+n)}$. Therefore, the parameter $n=1$ corresponds to the dipole case, $n=0$ corresponds to the split monopole case. $0<n<1$ corresponds to the twisted dipole case. The function $f(\theta) \approx 1-\cos^2\theta =\sin^2\theta$ is the angular dependence of the flux function. As have been shown in previous works (Pavan et al. 2009; Tong 2019), the exact function of $f(\theta)$ is similar to the dipole case for small twist. This is especially true for $\theta =0$ (pole) or $\theta=\pi/2$ (equator), where the difference is smaller. Similar things also happen in the black hole case (Tchekhovskoy et al. 2008), where the angular dependence can be approximated by the split monopole case.

Equation (\ref{eqn_Psi_main}) and (\ref{eqn_Phi_main}) are the approximate analytical solution to the pulsar equation, i.e., for  a rotating twisted magnetosphere. A straightforward way to check this point is to substitute these two equations into the original pulsar equation. Considering previous experiences (Narayan et al. 2007; Tchekhovskoy et al. 2008; Chen \& Zhang 2021), the approximation is rather good for small $\theta$ (i.e., the polar cap region). For other values of $\theta$, the same expression can only be viewed as approximations. The calculation is straightforward\footnote{The blackhole case mainly focus on the case of flux increasing with radius: $\Psi = r^{\nu} f(\theta)$, where $0\le \nu \le 2$. Considering that $\nu$ is just a constant parameter, the calculations in the blackhole case are still valid at present. Except for this point (a different radial dependence), previous blackhole studies have already confirmed that eq.(\ref{eqn_Psi_main}) and (\ref{eqn_Phi_main}) are the approximate solutions to the pulsar equation.}.

The basic picture is that (similar to that in the numerical simulations, Ntotsikas et al. 2024, 2025): (1) In the closed field line region, the poloidal field  is describe by the twisted dipole field (eq.\ref{eqn_Psi_main}). The toroidal field there is due to twist of the field, already fixed by the parameter $n$ for the self-similar case. (2) In the open field line regions, the poloidal field is also described by the flux function (eq.\ref{eqn_Psi_main}). The toroidal field is in the open field line regions is due to rotation of the neutron star (eq.\ref{eqn_Phi_main}). There are many assumptions and approximations in doing so. We will explore the consequences of this solution and try to compare it with the up-to-date numerical solutions (Ntotsikas et al. 2024, 2025). In this way, we hope this approximate analytical solution can be justified.

\subsection{The Y-point radius}

Assuming the approximate analytical solution, from the definition of the magnetic field, the three components of the magnetic fields are (for open field line regions):
\begin{eqnarray}
  B_r &=& \frac{1}{r^2 \sin\theta} \frac{\partial \Psi}{\partial \theta} = r^{-(2+n)} \frac{1}{\sin\theta} \frac{\partial f}{\partial \theta} \\
  B_\theta &=& -\frac{1}{r\sin\theta} \frac{\partial \Psi}{\partial r} = n r^{-(2+n)} \frac{1}{\sin\theta} f(\theta) \\
  B_\phi &=& \frac{1}{r\sin\theta} \Phi = -2 r^{-(2+n)} \frac{r}{R_{\rm lc}} \frac{1}{\sin\theta} f(\theta).
\end{eqnarray}
The main quantities related to the function $f(\theta)$ is: $f(\theta= \pi/2) =1$ (conservation of  flux), $f^\prime (\theta=\pi/2) =0$ (symmetry about the equator). If needed, the approximation for $f(\theta)$ can be used: $f(\theta) \approx \sin^2\theta$. For a neutron star with surface polar magnetic field $B_p$, the correspond flux function should include a normalization factor $(1/2) B_p R^{2+n}$, where $R$ is the neutron star radius. Then at the surface of the neutron star, the flux function is: $\Psi (r=R)= (1/2)B_p R^3 f(\theta)/R$. Neglecting a normalization factor $(1/2)B_p R^3$, the flux function at the surface is: $\Psi(r=R) = f(\theta)/R$. This expression is independent on the factor $n$, and is valid for both the dipole and twisted dipole case.

We are mainly interested in the physics of open field line regions, especially along the last open field line.  The ratio between the toroidal field and the poloidal field along the last open field line is (at the equator region, neglecting the numerical factors):
\begin{equation}
  \frac{B_\phi}{B_p} \sim \frac{r}{R_{\rm lc}}.
\end{equation}
Therefore, at the light cylinder radius $r\sim R_{\rm lc}$, the toroidal field and the poloidal field are comparable. This is the general concept of light cylinder for normal pulsars (Contopoulos et al. 1999; Harding et al. 1999). The exact relation between the toroidal field and the poloidal field may involve an arbitrary factor.

At present, we neglect this kind of factor and simply assume that the Y-point radius is at where the toroidal field and poloidal field are equal $B_p \sim B_\phi$. At the equator, the poloidal field is dominated by $B_\theta$. Therefore, the relation $B_p \sim B_\phi$ becomes:
\begin{equation}
  n r^{-(2+n)} \sim 2 r^{-(2+n)} \frac{r}{R_{\rm lc}}.
\end{equation}
The Y-point radius can be obtained:
\begin{equation}\label{eqn_RY_initial}
  R_Y \sim \frac{n}{2} R_{\rm lc}.
\end{equation}
Therefore, for a dipole field with $n=1$, the Y-point radius is $0.5$ times the traditional light cylinder radius. This may seem ridiculous at first sight. However, according to Contopoulos et al. (2024a,b), the Y-point radius may lie at $0.8-0.9$ the light cylinder radius (see also particle-in-cell simulations, Philippov et al. 2015). In our opinion, there is an undetermined numerical factor in eq.(\ref{eqn_RY_initial}). We require that for a dipole field, the Y-point radius equals the light cylinder radius. This will fix the numerical factor in the above express. 
And the final relation between the Y-point radius and the light cylinder radius is\footnote{This equation can be easily extended to a form: $R_Y = y n R_{\rm lc}$, where $y$ may be in the range $(0.8-0.9)$ considering recent numerical results for a dipole field ($n=1$). The modifications to the following results are only quantitative.}:
\begin{equation}\label{eqn_RY}
  R_Y = n R_{\rm lc}.
\end{equation}
For a twisted dipole field $0<n<1$, the Y-point radius will be smaller than the light cylinder radius. This is the major finding of our approximate analytical solution and will be the foundation of subsequent calculations.

\subsection{The polar cap}

A constant flux function corresponds to the projection of field lines on the $r-\theta$ plane:
\begin{equation}
  \Psi = r^{-n} f(\theta) = \mbox{const}.
\end{equation}
Along the last open field line, at $\theta =\pi/2$, $f(\theta=\pi/2) =1$. The radial extent reaches its maximum $r=R_Y$. At the surface of the neutron, the foot point of the field line corresponds to the polar cap:
\begin{equation}
  \frac{1}{R_Y^n} = \frac{f(\theta_{\rm pc})}{R^n},
\end{equation}
where $R$ is the neutron star radius, $\theta_{\rm pc}$ is the polar cap angular radius. Therefore, the angular flux function is:
\begin{equation}
  f(\theta_{\rm pc}) = \left( \frac{R}{R_Y} \right)^n.
\end{equation}
Using the approximation for $f(\theta) \approx \sin^2\theta$, the polar cap radius for a rotating twisted magnetosphere is:
\begin{equation}\label{eqn_thetapc}
  \sin\theta_{\rm pc} = \left( \frac{R}{R_Y} \right)^{n/2} = \left( \frac{R}{n R_{\rm lc}} \right)^{n/2}.
\end{equation}

This can be compared with eq.(12) in Tong (2019), where the polar cap for a twisted magnetosphere is obtained by introducing the light cylinder radius: $\sin\theta_{\rm pc} \approx (R/R_{\rm lc})^{n/2}$. The present result (eq.(\ref{eqn_thetapc})) differs from the that in Tong (2019) by a numerical factor in the denominator. Therefore, the previous conclusions in Tong (2019) are still valid (untwisting of a globally twisted magnetosphere due to physics in the open field line regions). The changes are only quantitative. However, Tong (2019) introduced the light cylinder radius as the starting point. Therefore, the Y-point radius can not be obtained self-consistently. While here the Y-point radius (eq.\ref{eqn_RY}) is obtained using the approximate analytical solution to the pulsar equation. And the polar cap radius is obtained in a self-consistent way.

\subsection{The particle outflow luminosity}

The work done on the outflowing current from the two polar caps is (Gruzinov 2005; Timokhin 2006):
\begin{equation}
  L = 2\int_0^{\Psi_0} \Phi d\Psi \propto \Psi_0^2,
\end{equation}
where $\Psi_0$ is the flux associated with the last open field line. From its definition, $\Phi$ denotes the poloidal current (up to a numerical factor). Using the solution for $\Phi$ (eq.\ref{eqn_Phi_main}), the particle outflow luminosity is proportional to $\Psi_0^2$. It can be shown that the above definition is equal to the Poynting flux in the open field line regions (Stefanou et al. 2023). Another point of view is: this definition is actually the work done on the outflowing particles. The current is proportional to $\Phi$, the maximum acceleration potential is proportional to $\Psi$. Therefore, the power defined in this way is just the current times the voltage. Therefore, we'd like to call it the ``particle outflow luminosity". Of course, here the ``particle outflow" means both the outflowing particles and the electromagnetic fields associated with them.

There is a tricky point here. In the case of normal pulsars, the rotational energy is the only energy source. Therefore, the particle outflow luminosity can be named as the rotational energy loss rate. In the case of magnetars, the magnetic energy release may dominate over the rotational energy loss rate, e.g. magnetar's persistent X-ray luminosity can be higher than their rotational energy loss  rate. In this case, the particle outflow luminosity may also be powered by the magnetic energy release. Noting this point, we just call it particle outflow luminosity.

The flux function is proportional to $r^{-n} f(\theta)$ (eq.\ref{eqn_Psi_main}). Of course, there are some numerical factors omitted (e.g., the polar magnetic field strength etc). Then the open field line flux will be $\Psi_0 \propto f(\theta_{\rm pc}) \propto (R/R_Y)^n = (R/n R_{\rm lc})^n$. For a rotating twisted magnetosphere, the particle outflow luminosity normalized to the dipole case is:
\begin{equation}\label{eqn_L}
  \frac{L_{\rm twist}}{L_{\rm dipole}} = \frac{(R/n R_{\rm lc})^{2n}}{(R/n R_{\rm lc})^{2n}|_{n=1}}
  = \left( \frac{1}{n} \right)^{2n} \left( \frac{R_{\rm lc}}{R} \right)^{2-2n}.
\end{equation}
Using the approximation for $f(\theta) \approx \sin^2\theta$, the particle outflow luminosity will be $L \propto \sin^4\theta_{\rm pc}$. This is similar to eq.(19) in Tong (2019), where the particle outflow luminosity is obtained in a different way (assuming a Goldreich-Julian current and maximum acceleration potential for the outflowing particles).

From the above express, it can be seen that the particle outflow luminosity for a rotating twisted magnetosphere can be much larger than the rotating dipole case. This may explain the enhanced spin-down during magnetar outburst (Archibald et al. 2020; Huang et al. 2023). Furthermore, during the outburst, the magnetosphere may evolve towards the dipole case, i.e. $n$ evolves from some initial value $n_0$ towards $n=1$. Then the spin-down torque will decrease with time during the outburst. This is consistent with the general trend of spin-down evolution during magnetar outburst.

Here the twist of the field line is characterized by the parameter $n$. Geometrically, the twist of the field line from the north pole to the south pole is defined as the maximum twist $\Delta \phi_{\rm max}$ (Thompson et al. 2002). There is a one-to-one relation between the parameter $n$ and the maximum twist (Pavan et al. 2009; Tong 2019):
\begin{equation}\label{eqn_phi_max_self-similar}
  \Delta \phi_{\rm max} = 2 \times \sqrt{\frac{35}{16} (1-n)}.
\end{equation}
The maximum allowed twist is $\pi$ radians for a twisted dipole field (Lynden-Bell \& Boily 1994; Wolfson 1995; Thompson
et al. 2002). The analytical result of the self-similar solution (eq.\ref{eqn_phi_max_self-similar}) can catch this point approximately. And the analytical result agrees with the detailed calculations, especially for small twist case (Pavan et al. 2009; Tong 2019).
During the numerical simulations (Ntotsikas et al. 2024, 2025), the twist defined there is half the value reported in previous works. Therefore, in order to compare  with the numerical simulations, half the maximum twist should be used:
\begin{equation}\label{eqn_Delta_phi}
  \Delta \phi \equiv \Delta \phi_{\rm max}/2 = \sqrt{\frac{35}{16} (1-n)}.
\end{equation}
Then the Y-point radius, polar cap, and particle outflow luminosity can all be obtained as a function of twist $\Delta \phi$.

\section{Comparison with the numerical simulations}

\begin{figure}
  \centering
  \includegraphics[width=0.45\textwidth]{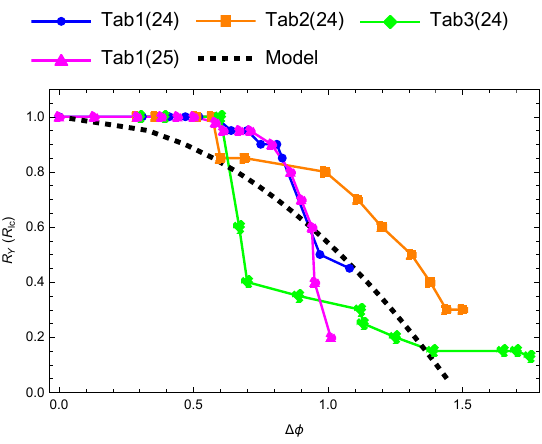}\\
  \caption{Y-point radius vs. the twist. The Y-point radius is in units of the light cylinder radius. The blue dots, orange squares, and green diamonds are from Table 1 to 3 in Ntotsikas et al. (2024), respectively. The magenta triangle is from Table 1 in Ntotsikas \& Gourgouliatos (2025). The black dashed line is the analytical approximation, eq.(\ref{eqn_RY}).}\label{fig_RY}
\end{figure}

\begin{figure}
  \centering
  \includegraphics[width=0.45\textwidth]{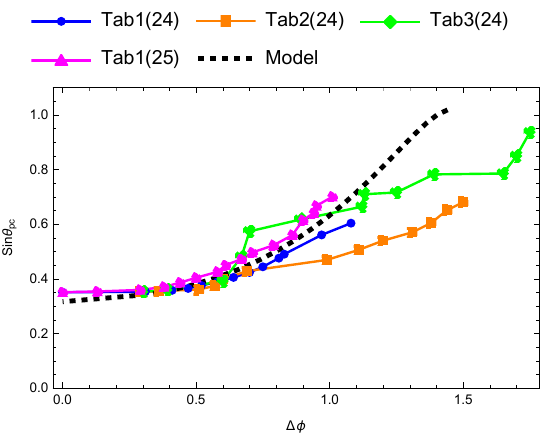}\\
  \caption{Polar cap angular radius vs. the twist. The analytical modeling is from eq.(\ref{eqn_thetapc}). The labeling is similar to figure 1.}\label{fig_thetapc}
\end{figure}

\begin{figure}
  \centering
  \includegraphics[width=0.45\textwidth]{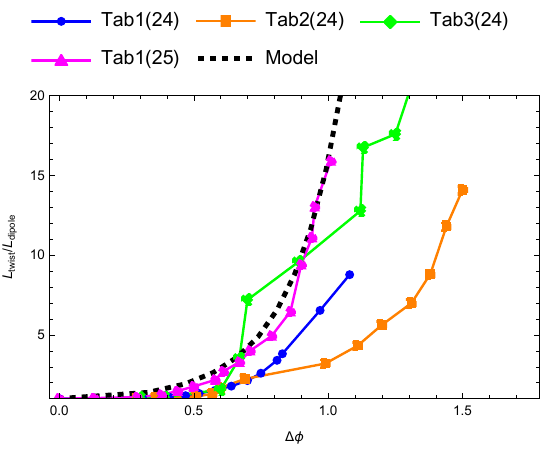}\\
  \caption{Particle outflow luminosity vs. the twist. The particle outflow luminosity is normalized to that of the dipole case. The analytical modeling is from eq.(\ref{eqn_L}). The labeling is similar to figure 1.}\label{fig_L}
\end{figure}

Previous numerical simulations have not include the twist of the magnetic field (Contopoulos et al. 1999; Spitkovsky 2006). Gruzinov (2006) had discussed that adding twist to the field lines may result in the opening of more field lines. In Ntotsikas et al. (2024, 2025), the twist in the closed field line regions is included, while the twist in the open field line regions is still due to the rotation of the neutron star. In the numerical simulations, it is found that (Ntotsikas et al. 2024, 2025): (1) The Y-point radius is smaller than the light cylinder radius, (2) The polar cap is larger compared with the dipole case, (3) the particle outflow luminosity is higher. Point (2) \& (3) are already known in the analytical result in Tong (2019). Here, all three points are obtained analytically (see the above section). The analytical solutions will be compared with the numerical simulations.

In the numerical simulations, the toroidal field (or poloidal current) in the closed field lines are prescribed in a power-law form. While, in our approximate analytical solutions, the closed field line region is assumed to be in a self-similar form. Despite this difference, the maximum twist can be defined in both cases. Therefore, we will compare the Y-point radius, polar cap, and particle outflow luminosity as a function of the maximum twist. Of course, the twist reported in the numerical simulations is actually half the maximum value.

In table 1-3 in Ntotsikas et al. (2024), table 1 in Ntotsikas \& Gourgouliatos (2025), the numerical results of the flux associated with the last open field line $\Psi_0$, normalized particle outflow luminosity compared with the dipole case $L_{\rm twisted}/L_{\rm dipole}$, Y-point radius $R_Y$ and half the maximum twist $\Delta \phi$ are shown. In the presence of twist, the field line will be different from the dipole case, the polar cap should be found by tracing back the field line to the neutron star surface (Ntotsikas \& Gourgouliatos 2025). As an approximation, the flux function will be like the dipole case at the neutron star surface (eq.(19) \& (20) in Ntotsikas et al. 2024. Actually, it is one of the boundary conditions in the numerical simulations. This expression holds for both the dipole and twisted dipole case, as has been proved above.):
\begin{equation}\label{eqn_psi0}
  \Psi_0 = \frac{\sin^2\theta_{\rm pc}}{R}.
\end{equation}
Therefore, the polar cap can be obtained from the result of $\Psi_0$:
\begin{equation}
  \sin\theta_{\rm pc} =\sqrt{\Psi_0 R}.
\end{equation}
This approximation is only used to convert the numerical value of $\Psi_0$ to that of $\sin\theta_{\rm pc}$.

The comparisons between our approximate analytical solutions and the numerical simulations are shown in figure \ref{fig_RY}, \ref{fig_thetapc}, and \ref{fig_L}. From this comparison, it can be seen that: (1) the general trend of both the analytical approximations and the numerical results are consistent with each other. (2) For the particle outflow luminosity, the analytical results and the numerical simulations of Ntotsikas \& Gourgouliatos (2025) are consistent with each other (magenta triangle in figure \ref{fig_L}). We consider these as two pieces of evidence in support of our approximate analytical solutions.

\section{Discussions}

\begin{figure}
  \centering
  \includegraphics[width=0.45\textwidth]{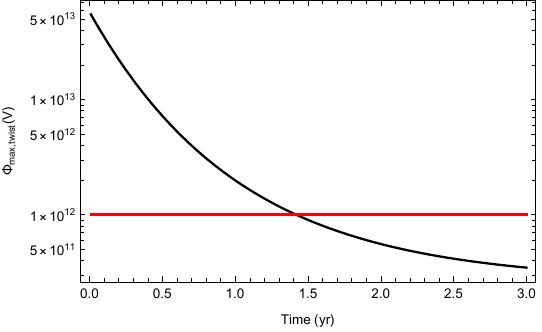}\\
  \caption{Maximum acceleration potential for an evolving twisted magnetosphere. The typical threshold potential for radio generation is about $10^{12} \ \rm V$. }\label{fig_Vmax}
\end{figure}

\begin{figure}
  \centering
  \includegraphics[width=0.45\textwidth]{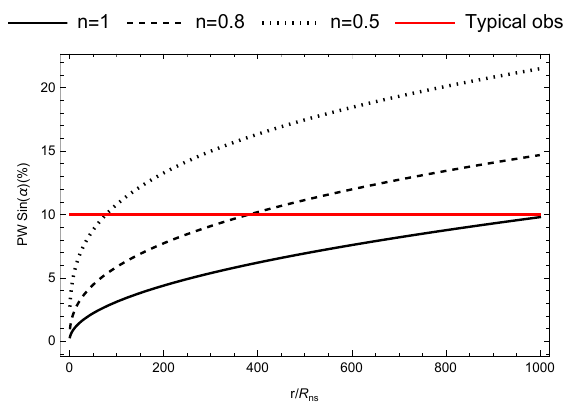}\\
  \caption{Pulse width as a function of emission height, for different values of $n$. The emission height is in units of the neutron star radius. The pulse width is multiplied by $\sin\alpha$, the unknown inclination angle. A typical pulse width of magnetar radio emission about $10 \ \%$ is shown.}\label{fig_PW}
\end{figure}

From the magnetohydrodynamics point of view, the magnetic field structure (eq.\ref{eqn_Psi_main} and \ref{eqn_Phi_main}) forms the basis of all subsequent calculations. Equation (\ref{eqn_Psi_main}) is the result of self-similar twisted dipole field. In the closed field line region, the relation between the current and flux is non-linear (Wolfson 1995). While, eq.(\ref{eqn_Phi_main}) represents a linear relation between the current and flux in the open field line regions, and it agrees with the minimum torque model. In general, the relation between the current and flux can be non-linear. The poloidal field is unchanged. Only the toroidal field changes for a different current. (1) For a split monopole solution (Contopoulos 2005): $\Phi = -\Omega \Psi (2-\Psi/\Psi_{\rm open})$, it is consistent with eq.(\ref{eqn_Phi_main}) for small $\theta$ (near the magnetic pole). For the toroidal field at the Y-point ($\Psi = \Psi_{\rm open}$), it is different from eq.(\ref{eqn_Phi_main}) by a factor of two. Since there is an undetermined numerical factor during the definition of the Y-point radius, the final express for the Y-point radius will be the same (eq.(\ref{eqn_RY})). (2)For other non-linear forms between the current and flux, the toroidal field, Y-point radius etc may be calculated similarly. However,
\begin{enumerate}
  \item we are not sure whether a non-linear relation between the current and flux is still an approximate solution to the original pulsar equation.
  \item Irrespective of the details, from previous experiences of analytical and numerical calculations (Tong 2019; Ntotsikas et al. 2024), the general trend should be the same: more twist  will result in more open field lines, and consequently larger polar cap and particle outflow luminosity.
\end{enumerate}

Here we want to discuss the radio emission of magnetars. The fast radio bursts may also be related with the magnetospheric physics of magnetars. The polar cap (eq.\ref{eqn_thetapc}) is related directly to the observed hot spot during magnetar outburst. During the outburst, an untwisting magnetosphere will result in a shrinking hot spot (for the parameter $n$ evolves from $n_0$ to 1). At the same time, the maximum acceleration potential across the polar cap is closed related to the turn-on/off of radio emission (Ruderman \& Sutherland 1975):
\begin{equation}
  V_{\rm max} = \frac{\Omega B_p R^2}{2c} \sin^2\theta_{\rm pc}.
\end{equation}
For a given parameter $0<n<1$, the corresponding maximum acceleration potential can be obtained. In order to obtain the evolution of magnetosphere, the evolution of the twist should be calculated (i.e., the evolution of $n$). This will depend on the particle acceleration mechanism (Beloborodov 2009; Tong 2019). At the same time, the magnetar X-ray flux may decay with time in an exponential form in observations (Coti Zelati et al. 2018). Therefore, the magnetic free energy stored in the twisted magnetic field may also decay with time in an exponential form. Assuming this exponential form of evolution, a simplified solution can be found for the parameter $n$ (Tong \& Huang 2020):
\begin{equation}\label{eqn_n}
  n(t) = 1- (1-n_0)e^{-t/\tau},
\end{equation}
where $n_0$ is the initial value of $n$, $\tau$ is the decaying timescale.

Using this approximation, a typical evolution of the maximum acceleration potential of a magnetar can be obtained, in figure \ref{fig_Vmax}. During the calculations, a typical period of $5\ \rm s$ and polar magnetic field strength $B_p =10^{12} \ \rm G$ is assumed. The initial value and decaying timescale of $n$ is: $n_0=0.5$, $\tau=1 \ \rm yr$. Figure \ref{fig_Vmax} demonstrate that if a magnetar's true dipole field is only about $10^{12} \ \rm G$ (its spin-down rate may be greatly enhanced by the particle outflow luminosity, therefore its true magnetic field may be much smaller than the characteristic magnetic field), during the outburst its polar cap and maximum acceleration potential may be greatly enhanced. The maximum acceleration potential may exceed $10^{12} \ \rm V$, the threshold value for typical radio emission. During the outburst decaying phase, its acceleration potential decays with time. Later, it may even fall below the radio emission threshold. Then the radio emission may turn off. This may be the first reason for the radio turn-on/off of magnetars.

From eq.(\ref{eqn_thetapc}), at radius $r$, the magnetic field opening angle is will be $(r/n R_{\rm lc})^{n/2}$. Similar to normal radio pulsar, the pulse width can be obtained for a given emission height $r$ and geometry (mainly the inclination angle $\alpha$, when the impact angle $\beta$ is assumed to be small) (Tong 2023):
\begin{equation}
  PW = \frac{3}{2\pi} \sin^{-1} \left( \frac{r}{n R_{\rm lc}} \right)^{n/2} \frac{1}{\sin\alpha}.
\end{equation}
This definition of pulse width is in the range $0-1$. Therefore, it is actually the duty cycle of the pulse profile. Then for a typical magnetar with period $P=5 \ \rm s$, the pulse width as a function of emission height for several values of $n$ are given in figure \ref{fig_PW}. For a larger twist (smaller $n$), the pulse width will be larger. Compared with normal pulsars, the radio emission of magnetar typically have wide pulse profile, by noting that magnetar have a long pulsation period (Camilo et al. 2006, 2008; Levin et al. 2012; Huang et al. 2021). From figure \ref{fig_PW}, for a typical pulse width about $10\%$ of magnetars, (1) a very high emission height is required for a dipole field ($n=1$), (2) for a twisted magnetosphere, the pulse width can  be explained by an emission height about 100 neutron star radius (the typical value of normal pulsars, Johnston \& Kramer 2019). During the outburst decaying phase, the pulse width decreases with time. It is possible that at a later time, the line of sight will be out-of the radio emission beam. This may be the second reason for the radio turn-on/off of magnetars.

A general trend of magnetar outburst can be obtained. For an evolving magnetosphere (parameterized by an increasing $n(t)$): (1) The Y-point radius will increase with time, until it reaches the light cylinder (the dipole case). (2) The polar cap will be a decreasing function of time. This may corresponds to a shrinking hot spot of magnetar X-ray ray emission. At the same time, a shrinking polar cap will also result in a decreasing maximum acceleration potential, and decreasing radio emission beam. (3) The particle outflow luminosity will also decrease with time. This may be responsible for the decreasing spin-down rate during magnetar outburst. By employing the simplified form (eq.\ref{eqn_n}) or some physical model, the time evolution of these three aspects can be calculated.

\section{Conclusions}

We have tried to provide an approximate analytical solution for the rotating twisted magnetosphere of magnetars. The poloidal field is approximated by the self-similar twisted dipole field (eq.\ref{eqn_Psi_main}). The toroidal field is approximated by the minimum torque model (eq.\ref{eqn_Phi_main}). Starting from these two equations, the Y-point radius (eq.\ref{eqn_RY}), polar cap (eq.\ref{eqn_thetapc}), and particle outflow luminosity (eq.\ref{eqn_L}) for a rotating twisted magnetosphere are obtained.

Comparisons with the numerical simulations are made. The general trend of the approximate analytical solution and the numerical simulations are the same. This may be viewed as support for the analytical approximation. The analytical solutions can be used to model the evolution of the magnetar magnetosphere. We have only showed the case of evolving maximum acceleration potential, and pulse width. These two points may be related to the radio turn-on/off of magnetars. In our opinion, the modeling of radio emission in the magnetar magnetosphere (magnetars and fast radio bursts) should be based on the rotating twisted magnetosphere, instead of the normal dipole case. In general, a rotating twisted magnetosphere has larger open field line regions. The radio emission of magnetars and fast radio bursts may both originate in the larger and evolving open field line regions of magnetars.

As can be seen from figure \ref{fig_RY}-\ref{fig_L}, the grid of the present numerical simulations are rather crude. We hope that future more refined numerical simulations will provide a more smooth curve.
A more detailed comparison between the approximate analytical solutions and numerical simulations can be made then.

\section*{Acknowledgements}

This work is supported by NSFC (12133004, 12173066) and National SKA Program of China (No. 2020SKA0120300, 2022SKA0120102).

\section*{Data availability}

This is a theoretical paper, mainly analytical. All the formulae are available in the article.








\bsp	
\label{lastpage}
\end{document}